\documentclass[prb,showpacs,twocolumn]{revtex4-1}
\usepackage{graphicx}
\usepackage{color}
\usepackage{booktabs}
\usepackage{array}
\usepackage{amssymb,amsmath}

\begin{document}
\title{Impurities as a source of flicker noise in graphene}
\author{A.~A.~Kaverzin$^{\dag,*}$}
\author{A.~S.~Mayorov$^\ddag$}
\author{A.~Shytov$^\dag$}
\author{D.~W.~Horsell$^\dag$}
\affiliation{\dag School of Physics, University of Exeter, Exeter, EX4 4QL, UK}
\affiliation{\ddag School of Physics and Astronomy, University of Manchester, Manchester M13 9PL, UK}

\begin{abstract}
We experimentally study the effect of different scattering potentials on the flicker noise observed in graphene devices on silica substrates. The noise in nominally identical devices is seen to behave in two distinct ways as a function of carrier concentration, changing either monotonically or nonmonotonically. We attribute this to the interplay between long- and short-range scattering mechanisms. Water is found to significantly enhance the noise magnitude and change the type of the noise behaviour. By using a simple model, we show that water is a source of long-range scattering.
\end{abstract}

\maketitle

The phenomenon of flicker noise (also known as $1/f$ noise) has been intensively studied in semiconductor structures \cite{Kogan,Kirton}. In Si MOSFETs it is ascribed to the random tunneling of electrons between the conducting channel and nearby impurity states, and measurements of the noise provide information about such states and their effect on the conduction \cite{Kirton}. A promising material to supercede silicon in future nanoelectronics is graphene \cite{Geim2009}. Electrical conduction through graphene is limited by various scattering mechanisms, which cannot be distinguished by their effect on the electronic transport. For instance, the linear relation between graphene conductivity and carrier concentration was initially attributed to scattering by Coulomb impurities \cite{Ando2006}. Later, it was realised that other scatterers, such as vacancies and ripples, produce nearly identical dependences \cite{Stauber2007,Katsnelson2008}. Recent measurements of flicker noise in monolayer graphene have shown that this noise is sensitive to the method of fabrication of the device \cite{AvourisNL,Wang2010,Ghosh}, which suggests that it results from several distinct sources of scattering. In this work, we measure the flicker noise in graphene on top of a SiO$_2$ substrate and demonstrate that it is possible to use such measurements as a sensitive tool to distinguish between short- and long-range scattering mechanisms. We identify water molecules as a source of long-range scattering and show that their removal by thermal annealing has a dramatic effect on the flicker noise.

Graphene transistors were prepared by mechanical exfoliation of graphene flakes onto $n^+$ Si/SiO$_2$ wafers with oxide thickness $300\,$nm. Multiple Au/Cr contacts were made to each flake. The flakes had dimensions ranging from $1$ to $4\,\mu$m in width and from $5$ to $22\,\mu$m in length and were verified to be monolayers by means of Raman spectroscopy \cite{FerrariPRL97} and measurement of plateau positions in the quantum Hall regime \cite{NovoselovN}. The concentration of carriers was tuned by applying gate voltage, $V_G$, between the substrate and the graphene. Four-terminal measurements of the resistance were carried out at $300\,$K either in an inert He atmosphere or in vacuum. A total of $8$ samples, named S1 to S8, were studied in detail. Samples S4-S8 were measured both before and after annealing (designated here by an asterisk) at a temperature of $140\,^{\circ}$C for about $1$ hour.
\begin{figure}[htb]{}
\includegraphics[width=.98\columnwidth]{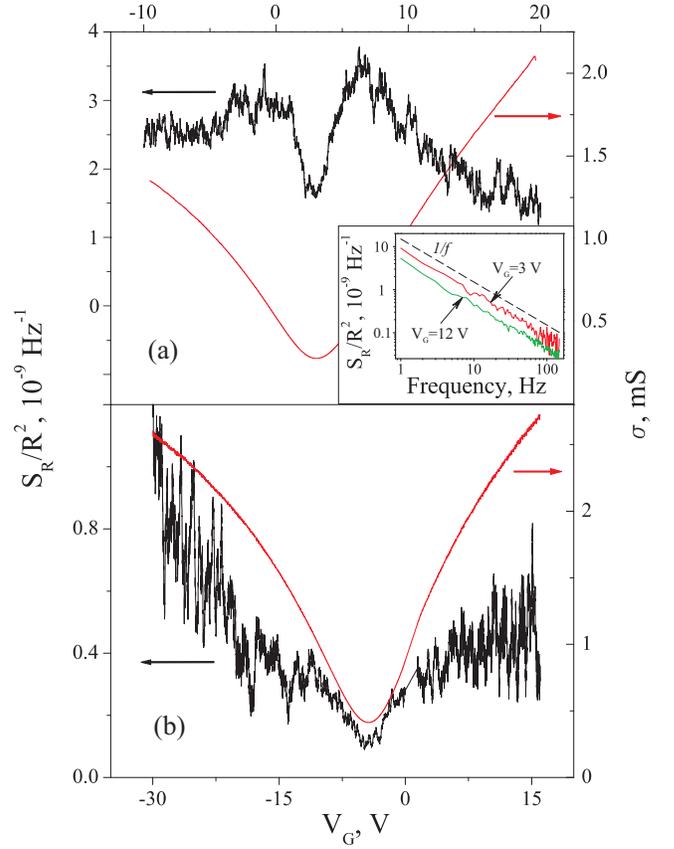}
\caption{(Color online) Normalised noise spectral density $S_R/R^2$ at $f=30\,$Hz (black) and conductivity $\sigma$ (red) are shown as functions of the gate voltage $V_G$. Two distinct behaviours of the noise can be seen: (a) $M$-type behaviour in sample S4 and (b) $V$-type behaviour in sample S5*. Inset: $S_R/R^2$ for sample S1 at $V_G=3\,$V (red) and $12\,$V (green) from the Dirac point. A theoretical $1/f$ noise spectrum is shown as a dashed line.}
\label{Fig2}
\end{figure}

The low-frequency noise ($\lesssim200\,$Hz) was measured using both a spectrum analyser and a lock-in amplifier. The spectrum analyser was used to check the $1/f$ scaling of the power spectral density with frequency. (A cross-correlation technique ensured that the noise contribution coming from the voltage amplifiers was minimised.) To avoid Joule heating of the sample and other nonlinear effects, the noise power spectra were confirmed to scale expectedly with the square of the applied source-drain bias voltages below $10\,$mV. The normalised noise power spectra $S_R/R^2$ of sample S1 are shown in the inset to Fig.~\ref{Fig2}. A clear $1/f^{\alpha}$ dependence, with $\alpha\simeq1$, was observed for all gate voltages in all samples.

Figures \ref{Fig2}(a) and \ref{Fig2}(b) show the dependence of normalised spectral density of the noise $S_R/R^2$ and conductivity $\sigma$ on $V_G$ for two samples, S4 and S5*, measured by a lock-in amplifier over a narrow frequency passband. These samples had similar mobilities, $\mu\simeq11000\,$cm$^2/$Vs, and were studied under similar experimental conditions (except that sample S5* was thermally annealed before measurements). In conventional semiconductor devices \cite{Kogan}, the low-frequency noise magnitude increases gradually with decreasing concentration $n$ in agreement with the Hooge relation. In our experiments, the noise exhibits the opposite behaviour: it experiences a minimum at the Dirac point and increases with $n$ at small concentrations. However, at larger concentrations the noise either continues to increase, as shown in Fig.~\ref{Fig2}(b), or reaches a maximum and decreases with $n$, Fig.~\ref{Fig2}(a). We shall refer to these two behaviours as $V$-type and $M$-type, respectively.

\begin{figure}[htb]{}
\includegraphics[width=0.98\columnwidth]{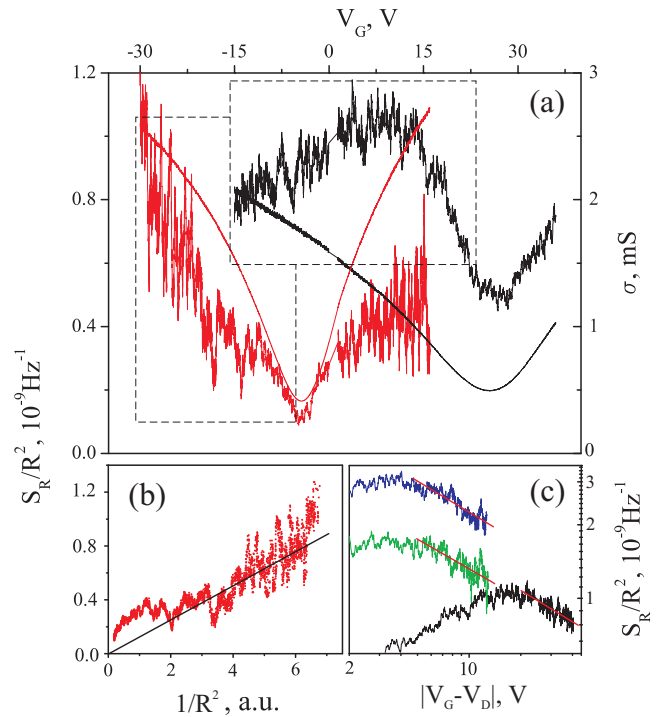}
\caption{(Color online) (a) Normalised noise, $S_R/R^2$, at $30\,$Hz and conductivity, $\sigma$, dependence on gate voltage before ($V_D\simeq24\,$V) and after ($V_D\simeq-4\,$V) annealing in sample S5. (b) The noise $S_R/R^2$ as a function of $1/R^2\propto\sigma^2$ for the same sample after annealing. The straight line shows the linear fit by Eq.~\eqref{ShR2}. (c) The noise dependences on the voltage from the Dirac point $|V_G-V_D|$ for three different samples in logarithmic scale. The fit $S_R/R^2\propto n^{-0.4}$ is shown as a straight line.}
\label{Fig3}
\end{figure}

It can be seen that the conductivity dependences on $V_G$ for the two samples are alike, while the noise curves are drastically different (although sample S5* exhibits a sublinear behaviour in $\sigma(V_G)$, which can be interpreted as a minor contribution from short-range scatterers \cite{Adam2009}). This suggests that the noise is more sensitive to the scattering mechanisms than the resistance is. Most of the samples exhibited $M$-type behaviour before annealing, which transformed into $V$-type after annealing, irrespective of the change in mobility, as shown in Fig.~\ref{Fig3}(a). One of the most important effects of annealing at temperatures above $100\,^\circ$C is the removal of atmospheric water from the sample surface, which adsorbs there during fabrication. Water is known to affect impurity scattering, as observed in measurements of conductance \cite{LohmannNL,LafkiotiNL,SchedinNatMat,Kaverzin2011}. Assuming that annealing mainly removes water impurities, we attribute the observed $M$-type dependence of the $1/f$ noise to water-like contaminants. The $V$-type dependence is, therefore, associated with impurities in the SiO$_2$ substrate and any impurities on the sample surface that cannot be removed by annealing at $140\,^\circ$C. In order to identify the scattering mechanism responsible for $M$-type dependence, we employ a simple theoretical model described below.

In the framework of Drude theory, the resistance of a graphene device $R$ is determined by carrier concentration $n$ and mobility $\mu$. Mobility, in turn, is a function of $n$ and the impurity concentration $N$. The $1/f$ noise resulting from fluctuations of $R(n,N)$ is, therefore, also determined by these two quantities. An example of a physical process in which both $n$ and $N$ are changing is charge trapping in the SiO$_2$. Electrons can tunnel between a trap and the conductive channel if the trap has an electron level close to the Fermi energy. These random tunneling events result in fluctuations of $n$ and $N$, which leads to the fluctuations of resistance
\begin{eqnarray}
\delta{R}=\frac{\partial{R}}{\partial{n}}\delta{n}+\frac{\partial{R}}{\partial{N}}\delta{N},\label{NMR2}
\end{eqnarray}
where $\delta{n}$ and $\delta{N}$ are fluctuations of $n$ and $N$, respectively. The noise spectral density $S_R$ is proportional to $(\delta R)^2$. It was shown that a uniform distribution of tunneling distances leads to an exponentially broad distribution of tunneling characteristic times, which results in $1/f$ scaling of $\delta n$ and $\delta N$ with frequency \cite{Hung1990}. The first term in Eq.~\eqref{NMR2} is associated with the fluctuations in the carrier concentration and can be experimentally obtained from the dependence of $R$ on $V_G$ using the relation between the gate voltage and concentration: $n=CV_G/e=V_G\cdot7.2\cdot10^{10}\,$cm$^{-2}/$V ($C$ is a flake-gate capacitance). Therefore, this term is proportional to $dR/dV_G$. The second term is determined by the scattering mechanism, which provides a particular dependence of $R(n,N)$, and will be considered later.

\begin{figure}[htb]{}
\includegraphics[width=.98\columnwidth]{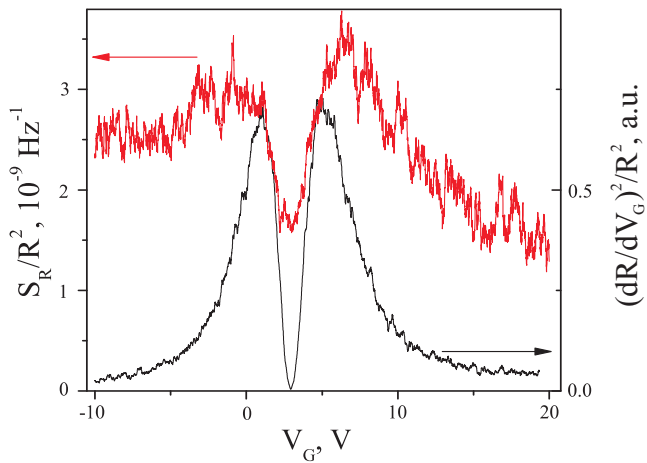}
\caption{(Color online) The normalised contribution from the first term in Eq.~\eqref{NMR2}, $(dR/dV_G)^2/R^2$ (black), and normalised noise, $S_R/R^2$ (red) are shown as functions of $V_G$ for sample S4.}
\label{Fig4}
\end{figure}

In Fig.~\ref{Fig4}, the normalised noise  $S_R/R^2$ and the contribution from the carrier concentration fluctuations, $(dR/dV_G)^2/R^2$, are plotted for sample S4. At first glance, the two quantities behave in a similar way. However, a closer inspection reveals two important differences. Firstly, the derivative vanishes at the Dirac point (a simple consequence of the resistance maximum), whereas the measured value of the noise is finite for all samples. Secondly, the maxima in $(dR/dV_G)^2/R^2$ occur at lower values of $V_G$ than those in the measured noise.

In our experiments, the maxima in $(dR/dV_G)^2/R^2$ always occur within $5\,$V of the Dirac point, whereas the $V$-type noise dependence extends well beyond this region. This implies a significant contribution from the second term in Eq.~\eqref{NMR2}, and can be understood if we assume that the mechanism causing resistance fluctuations in this case does not involve a change in $n$ but only in $N$. We show below that a likely source of such fluctuations is a scattering by short-range impurities.

Let us discuss the role of the second term in Eq.~\eqref{NMR2} considering two types of scatterers: short-range and long-range impurities \cite{Ando1998,Ando2006}, such as lattice defects and charged impurities. (Other scatterers, such as ripples, mid-gap states and phonons \cite{Stauber2007,Katsnelson2008,ChenSSC}, produce a mobility dependence on $n$ similar to that expected for short- and long-range scattering.) We will assume that these two mechanisms are independent of each other, and consider their contributions to resistance fluctuations separately. The contribution to the resistance of graphene due to a short-range scattering potential is concentration independent \cite{Ando1998}, while for a long-range potential it is inversely proportional to concentration \cite{Ando2006}.

Hypothetically, the resistance of the sample and its fluctuations can be determined by unrelated scattering mechanisms. A strong, but non-fluctuating, scattering potential can give rise to a substantial contribution to the resistance without having any effect on the noise. Similarly, a weak fluctuating potential may give a negligible contribution to the resistance, but be the dominant factor in the noise. Assuming that the main source of fluctuations is short-range impurities, while the resistance is determined by long-range impurities, i.e. $\delta R_{lr}{\ll}\delta R_{sr}$ and $R_{sr}{\ll}R_{lr}$, we find for the (normalised) spectral density of the noise that
\begin{eqnarray}
\frac{S_{R}}{R^2}\propto\left(\frac{\delta{R}}{R}\right)^2\sim\left(\frac{\delta{R_{sr}}}{R_{lr}(n)}\right)^2\propto{n}^2.\label{ShR}
\end{eqnarray}
In the opposite case, when long-range impurities are the source of noise and resistance is determined by long-range impurities,
\begin{eqnarray}
\frac{S_{R}}{R^2}\propto\frac{1}{n^2}. \label{LR}
\end{eqnarray}
Comparing these two extreme cases one can see that the details of impurity scattering determine the behaviour of the flicker noise at large carrier concentrations.

One can consider a more realistic case, when both types of impurities contribute to the resistance. If the main source of the fluctuations is short range impurities, similar analysis yields
\begin{eqnarray}
\frac{S_{R}}{R^2}\propto\frac{1}{R^2(V_G)}.
\label{ShR2}
\end{eqnarray}
We used this relation to fit the results for sample S4 after annealing for gate voltages between $-30$ V and $-10$ V (away from the Dirac point), Fig.~\ref{Fig3}(b). This qualitative agreement with Eq.~\eqref{ShR} and \eqref{ShR2} suggests that the $V$-type noise dependence is due to fluctuating short-range disorder.

The contribution to the noise from the carrier concentration fluctuations is mainly seen before annealing as the $M$-type dependence of the noise. The decrease of the noise with increasing concentration at large $V_G$ can be deduced either from the derivative $\partial{R}/\partial{n}$ or from Eq.~\eqref{LR}. Fitting this part of $M$-type dependence by $n^{-\alpha}$ gives us $\alpha\simeq0.4\pm0.1$, Fig.~\ref{Fig3}(c). This leads to the conclusion that the $M$-type dependence is determined by long-range scattering, although not in the extreme limit of Eq.~\eqref{LR}, and the main source of this scattering is contaminants left on the surface of the graphene flake after fabrication, such as atmospheric water.
\begin{figure}[htb]{}
\includegraphics[width=0.98\columnwidth]{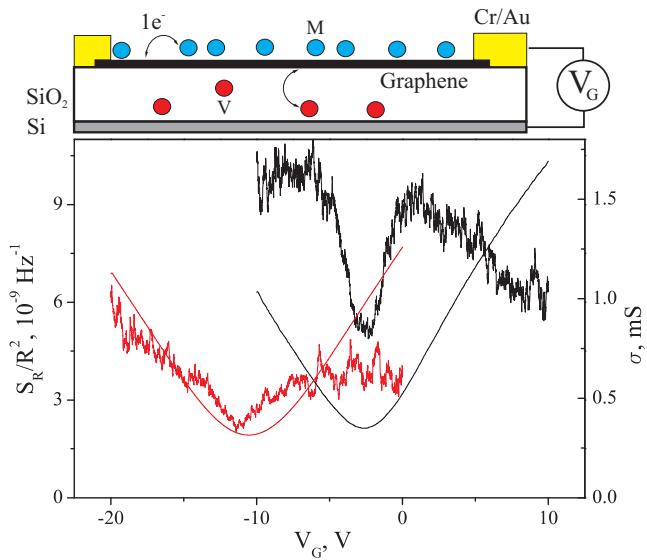}
\caption{(Color online) $S_R/R^2$ at $30\,$Hz and $\sigma$ dependence on gate voltage after doping by pure water ($V_D\simeq-3\,$V) and after annealing ($V_D\simeq-11\,$V) in sample S6. Noise after annealing is scaled up by factor of $4$. The top panel schematically shows tunneling events between graphene and impurities which cause both $M$-type (such as water) and $V$-type (short-range defects in SiO$_2$ or in the graphene itself) fluctuations.}
\label{Fig5}
\end{figure}

To test the hypothesis about the origin of $M$-type behaviour of the flicker noise, we doped sample S6 with pure water. To eliminate any possible contribution from air contaminants we first annealed the sample. Then, it was exposed to water vapour in an inert helium atmosphere and the flicker noise was measured. To revert the sample to its original state, it was then annealed again and the noise was measured. The results are shown in Fig.~\ref{Fig5}. The sample exhibits an $M$-type noise dependence after water doping, transforming to $V$-type after annealing. This demonstrates that it is possible to change reversibly the shape of noise dependence by this procedure. This also shows that water-like contaminants are a source of long-range scattering in graphene. It is interesting to note that the change in the noise magnitude due to annealing of water (Fig.~\ref{Fig5}) is more than tenfold, while the resistance is changed only by a factor of $2$ at all concentrations.

Finally, we look briefly at the noise behaviour around the Dirac point. Naively, in this region one would expect resistance fluctuations to be significantly suppressed since $R$ weakly depends on $n$. In all our samples, the noise is found to be finite and to increase with increasing carrier concentration, e.g. Figs.~\ref{Fig2} and \ref{Fig3}. The minimum, seen explicitly in the $M$-type dependence, indicates that the mechanisms controlling the noise are quite different from those at higher values of $n$. Near the Dirac point the Drude model is not applicable, and the noise cannot be described with the help of Eqs.~\eqref{ShR}-\eqref{ShR2}. Instead, the conductance is determined by an electron--hole puddle percolation network \cite{Jens} where local deviations of the potential create regions of finite electron or hole concentration across the graphene flake. It was shown that switching between different percolation patterns leads to the non-zero fluctuations of conductance even at the Dirac point \cite{Cheianov2007}, which can explain qualitatively the finite value of the noise we observe.

In conclusion, we have shown that flicker noise is a powerful tool for determining the scattering mechanisms in graphene. A distinctive behaviour of flicker noise for short-range and long-range scatterers was demonstrated. Water was shown to be a long-range scatterer and its presence on the graphene surface was found to increase the noise by an order of magnitude, yet cause a comparatively insignificant change in the resistance. Thus, we have demonstrated that low-frequency noise and resistance in graphene are determined by different scattering mechanisms.

We are grateful to A.K.Savchenko for inspiring this work, and to Roman Gorbachev, Freddie Withers and Reuben Puddy for help with sample fabrication. This research was supported by the EPSRC (Grant nos. EP/G036101/1 and EP/G041482/1).\\
\\
\noindent$^{*}$ak328@exeter.ac.uk

\bibliographystyle{apsrev}

\end{document}